\documentclass[aps, twocolumn, superscriptaddress, showpacs, nofootinbib, longbibliography]{revtex4-1}

\usepackage[utf8]{inputenc}
\usepackage[T1]{fontenc}
\usepackage{ae,aecompl} 
\usepackage{graphicx}
\usepackage{amsmath}
\usepackage{bm}
\usepackage{xcolor}
\usepackage{amssymb}
\usepackage{latexsym}
\usepackage{wasysym}
\usepackage{psfrag}
\usepackage{ifthen}
\usepackage{verbatim}
\usepackage[citecolor=blue,colorlinks=true]{hyperref}
\usepackage{longtable}
\usepackage{float}
\usepackage[utf8]{inputenc}
\usepackage{lineno}

\usepackage{aas_macros}
\usepackage[normalem]{ulem}
\usepackage{comment}
\usepackage{natbib}
\definecolor{bbsalmon}{rgb}{1.0, 0.47, 0.42}
\definecolor{datablue}{rgb}{0.0, 0.0, 1.0}

\begin{document}

\preprint{APS/123-QED}

\title{Gravitational wave memory of compact binary coalescence in the presence of matter effects }

\author{Dixeena Lopez}
\affiliation{%
 Physik-Institut, University of Zurich, Winterthurerstrasse 190, 8057 Zurich, Switzerland 
}%

\author{Shubhanshu Tiwari}%
\affiliation{%
 Physik-Institut, University of Zurich, Winterthurerstrasse 190, 8057 Zurich, Switzerland 
}%

\author{Michael Ebersold}
\affiliation{Laboratoire d'Annecy de Physique des Particules, CNRS, 9 Chemin de Bellevue, 74941 Annecy, France}

\begin{abstract}
Binary neutron stars (BNSs) and neutron star--black hole (NSBH) binaries are two of the most promising gravitational wave (GW) sources to probe matter effects. Upcoming observing runs of LIGO-Virgo-KAGRA detectors and future third generation detectors like Einstein Telescope and Cosmic Explorer will allow the extraction of detailed information on these matter effects from the GW signature of BNS and NSBH systems. One subtle effect which may be helpful to extract more information from the detection of compact binary systems is the nonlinear memory. In this work, we investigate the observational consequences of gravitational wave nonlinear memory in the presence of matter effects. We start by quantifying the impact of nonlinear memory on distinguishing BNS mergers from binary black holes (BBHs) or NSBH mergers. We find that for the third generation detectors, the addition of nonlinear memory to the GW signal model expands the parameter space where BNS signals become distinguishable from the BBH and NSBH signals. Using numerical relativity simulations, we also study the nonlinear memory generated from the postmerger phase of BNS systems. We find that it does not show a strong dependence on the equation of state of the NS. However, the amplitude of nonlinear memory from the BNS postmerger phase is much lower than the one from BBH systems of the same masses. Furthermore, we compute the detection prospects of nonlinear memory from the postmerger phase of NS systems by accumulating signal strength from a population of BNS mergers for the current and future detectors. Finally, we discuss the impact of possible linear memory from the dynamical ejecta of BNS and NSBH systems and its signal strength relative to the nonlinear memory. We find that linear memory almost always has a much weaker effect than nonlinear memory.




\end{abstract}

\maketitle

\section{Introduction} \label{intro}

The LIGO-Virgo-KAGRA collaborations have detected around 90 gravitational wave events, all coming from compact binary coalescence (CBC) mergers in all possible configurations of black holes (BHs) and neutron stars (NSs) \cite{GWTC3,ligo_2015,virgo_2014}. The majority of these events were identified as binary black hole mergers. Two were identified as binary neutron star mergers, namely GW170817 and GW190425 \cite{GW170817, GW190425}, and two as neutron star--black hole mergers, namely GW200105 and GW200115 \cite{NSBH-LVK}. Only one CBC event GW170817, the binary neutron star (BNS) merger event, was accompanied by the detection of an electromagnetic (EM) counterpart in the wide range of the EM spectrum \cite{GW170817-mult,GW170817-gam}. Also, only for GW170817, the identification of the event comes from not only the component mass measurements but also the measurement of tidal deformation and EM counterparts \cite{LIGOScientific:2018hze}. Two other noteworthy events are the GW signals GW190814 and GW200210 with an astrophysical probability of around 0.5, observed during the third observing run of the LIGO-Virgo detectors. They are compact binary merger events where the mass of the secondary companion has significant support in the mass range where both NS and BH can exist ($2.59^{+0.08}_{-0.09}M_{\odot}$ and $2.83^{+0.47}_{-0.42}M_{\odot}$ with 90\% credible intervals) \cite{GW190814,GWTC3}. Hence, ambiguity about the nature of the lighter compact object remains as the inference on the nature of the object comes only from the mass measurement. In the absence of an EM counterpart and tidal deformability measurement, the nature of the secondary companion (BH or NS) is not unambiguously known. To this end, the nonlinear memory can be of aid as was exemplified in the case of binary black holes (BBHs) and neutron star--black holes (NSBHs) as per the theoretical prediction \cite{NSBHmem_21}.


The GW memory effect describes the permanent displacement of an ideal GW detector---one that is only sensitive to gravitational forces---after a GW has passed. The memory appears in a linear and a nonlinear form. The linear memory is associated with unbounded sources of GW~\cite{Zeldovich-1974,braginskii-1987}, like hyperbolic orbits~\cite{turner_will-1978} and dynamical ejecta~\cite{kyutoku-2013}, whereas the nonlinear memory is produced by GWs themselves and therefore occurs in any system that radiates GWs~\cite{Christodoulou-1991,thorne-1992,favata-2008}. Although the final displacement is undetectable with interferometric detectors like LIGO and Virgo, the buildup of the memory over time adds a low-frequency component to GW signals that are potentially detectable. For a binary system, GW memory shows very slow growth in amplitude during the inspiral and a significant rise at the time of the merger; this is equivalent to a signal that is like a step function and after applying a low-frequency cut, will be a broadband GW burst. In this work, we only consider the displacement memory which is the leading order memory effect \cite{Mitman:2020pbt,N1,N2}.   

In Ref. \cite{NSBHmem_21}, the distinguishability of GW signals from NSBH and BBH systems with the inclusion of the GW memory waveform was discussed.
In this work, we extend this discussion to BNS systems. Since BNS systems can also have a postmerger stage that can produce memory \cite{yang-2018}, we also study the features and detectability of BNS postmerger memory. 

Since the memory signal always saturates at the low-frequency limit of the detector, it has a unique feature that allows probing sources and effects that are at higher frequency \cite{Ebersold:2020zah}, such as the BNS postmerger phase. However, the high noise level at lower frequencies of the detectors makes it challenging to measure memory from a single event for the current generation of detectors and hence stacking multiple events will be needed to detect memory \cite{lasky_GW150914mem, Review_Grant2022}. 

In this paper, we scrutinize the various effects of memory of compact binary systems that have the mass around the mass of neutron stars. We also study the prospects of the detectability of the memory from the BNS postmerger part in a population of BNS events for various equation of state (EOSs) using numerical relativity simulations for the current and future ground based detectors. The paper is organized as follows: Section~\ref{sect_2} describes the BNS, BBH, and NSBH waveform models considered to study the effect of GW memory. In this section, we also explain the intrinsic parameters of the CBC systems that contribute to nonlinear memory. In Sec.~\ref{sect_3}, we show the impact of GW memory on distinguishing different CBC systems. Section~\ref{sect_4} contains the results on prospects of detecting nonlinear memory from a population of BNS events. Section~\ref{sect_5} discusses the possibility of linear memory occurring from dynamical ejecta of BNS and NSBH systems interfering with the nonlinear memory. Finally, in Sec.~\ref{sect_6} we give our conclusion with some additional results shown in Appendixes ~\ref{app:coreDB},~\ref{app:nsbh} and~\ref{app:bbh_eff}.

\section{nonlinear memory waveform}\label{sect_2}


It is convenient to represent the plus and cross polarization of the gravitational waveform decomposed onto spin-weighted spherical harmonics. Assuming a nonspinning compact binary system orbiting in the $x-y$ plane with angular momentum along the $z-$ direction, the angles in the spin-weighted spherical harmonics are the inclination angle $\iota$, representing the angle between angular momentum and line of sight of the detector and $\phi^{\prime}$, representing the angle between the line of sight of the detector projected into the source plane and the $x-y$ axis:

\begin{equation}\label{eq:complex_wf}
h_{+}-i h_{\times}=\sum_{\ell=2}^{\infty} \sum_{m=-\ell}^{\ell} h_{\ell m}(t) {}_{-2}Y_{\ell m}\left(\iota, \phi^{\prime}\right) \,.
\end{equation}

In the case of quasicircular, aligned spin binaries, $h_{\ell m}-$ modes with $m \neq 0$ carry the ``oscillatory'' part, in which the 22-mode marks the dominant contribution. In the $m = 0$ modes, in particular $h_{20}$ and $h_{40}$, we find the GW memory which is a subdominant effect~\cite{mem_patricia}.
The nonlinear memory arises from GWs being scattered on already emitted GWs. Thus its contribution to the transverse-traceless (TT) piece of the metric perturbation can be computed from any GW energy flux 
 $dE_\mathrm{GW}/(d\Omega dt)$~\cite{wiseman-1991,favata-2008},
\begin{equation}\label{eq:mem_hjk}
h_{jk}^{\mathrm{TT, mem}} = \frac{4}{r} \int_{-\infty}^{t_r} dt \left[\int \frac{d E_\mathrm{GW}}{d \Omega^{\prime} dt} \frac{n_j^{\prime} n_k^{\prime}}{1-\mathbf{n}^{\prime} \cdot \mathbf{n}} d \Omega^{\prime}\right]^\mathrm{TT} \,,
\end{equation}
where $\mathbf{n}$ is the unit vector pointing from the source to the observer, $\mathbf{n^\prime}$ is a radial unit vector with $d \Omega^{\prime}$ being the solid angle element around the source, $r$ is the distance to the source, and $t_r$ denotes the retarded time. Using the polarization tensors $e_{+,\times}^{jk}$, the components of the metric perturbation can be cast on the GW polarizations $h_{+,\times} = h_{jk}^{\mathrm{TT}} e_{+,\times}^{jk}$, see e.g.~Ref.~\cite{favata-2008}. The GW energy flux, deducted from the GW stress-energy tensor can be written in terms of time derivatives of the waveform polarizations:
\begin{equation}\label{eq:GWflux}
\frac{dE_\mathrm{GW}}{dt\,d\Omega} = \frac{r^2}{16 \pi} \langle {\dot h_+}^2
 + {\dot h_\times}^2 \rangle \,,
\end{equation}
where the angled brackets mean to take an average over several wavelengths.
In order to perform the angular integration in Eq.~(\ref{eq:mem_hjk}), the GW energy flux is rewritten in terms of $h_{\ell m}$ modes and spin-weighted spherical harmonics by inserting Eq.~(\ref{eq:complex_wf}). The angular part can then be expressed elegantly in terms of Wigner-3-j symbols~\cite{Faye-2014}, thereby providing selection rules for which mode combinations contribute to memory. The memory generated by the oscillatory $h_{22}$ mode is found mainly in the  $h_{20}^{\mathrm{mem}}$ and subdominantly in the $h_{40}^{\mathrm{mem}}$ mode with plus and cross polarization given as
\begin{subequations}\label{eq:mem_plus}
\begin{align}
h^{\mathrm{mem}}_{+} &=  \frac{r}{192 \pi} \sin ^2 \iota\left(17+\cos ^2 \iota\right) \int_{-\infty}^{t_r} \left|\dot{h}_{22}\right|^2 dt \,,\\
h^{\mathrm{mem}}_{\times} &= 0 \,.
\end{align}
\end{subequations}

As apparent from Eq.~(\ref{eq:mem_plus}), the memory effect vanishes completely for face-on binary systems $(\iota = 0^\circ)$ and is maximal if the system is oriented edge-on $(\iota = 90^\circ)$. The GW memory from post merger phase is also expected to be dominant in the $h_{20}^{\mathrm{mem}}$ mode as the systems we have considered here have $h_{22}$ as the dominant mode for the oscillatory signal.


In this work, we consider the CBC systems where the matter effects are important; this happens when at least one companion is a NS. The GW waveform then has the imprints of matter effects which depend on the NS EOS primarily in two ways: tidal deformation during inspiral and postmerger oscillations of the remnant.   

 
\begin{figure*}[htp] 
  \centering
    {\includegraphics[width=1\linewidth]{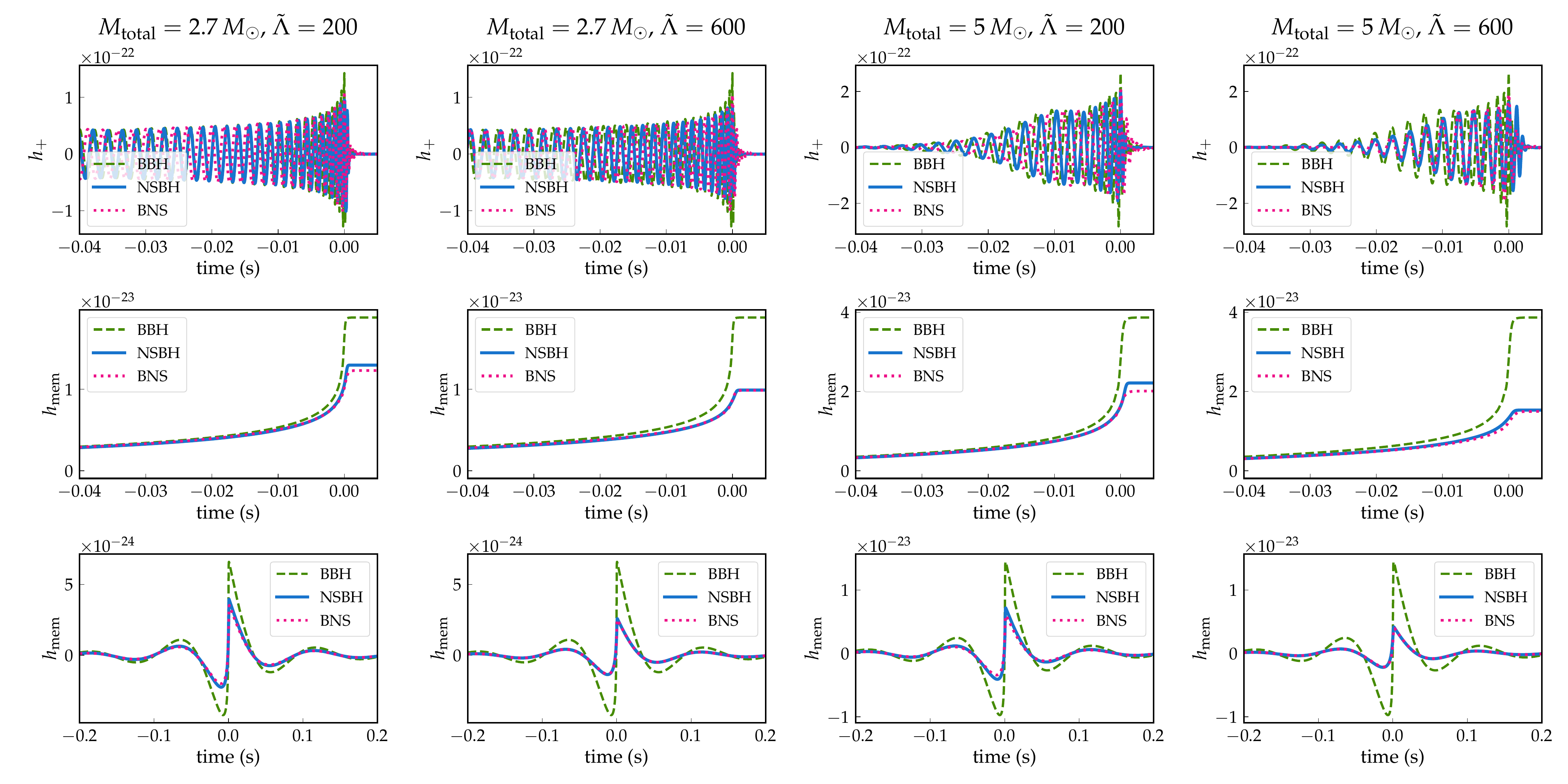}}
  \caption{The top panel plots \texttt{SEOBNRv4\_ROM\_NRTidalv2} oscillatory waveform for BNS and BBH, and $\mathtt{SEOBNRv4\_ROM\_NRTidalv2\_NSBH}$ for NSBH systems with equal mass binaries. The plots show four combinations of total mass and tidal deformability of the BNS system with $M_{\mathrm{total}}$ equal to $2.7M_{\odot}$ and $5M_{\odot}$ and $\Tilde{\Lambda}$ equal to 200 and 600. The NSBH system plot with $\Lambda_{\mathrm{NS}}$ equals 200 and 600. The second panel shows the corresponding nonlinear memory and the bottom panel plots the nonlinear memory waveform with a low-frequency cutoff of 10\,Hz. The distance to the source is fixed at 100 Mpc.}
  \label{fig.seobnrWF}
\end{figure*}


\subsubsection{Memory from the inspiral-merger part of the signal}

The inspiral part of the GW signal from BNS systems contains the corrections coming from the tidal deformability parametrized as a dimensionless quantity  $\Lambda$ defined as 
\begin{equation}
    \Lambda \equiv \frac{\lambda}{m^5} = \frac{2}{3} k_2 \frac{R^5}{m^5} \,,
\end{equation}
where $m$ and $R$ are the mass and radius of the star respectively, and $k_2$ is the Love number \cite{Katerina_tidal}.
In the case of NSs, $k_2$ typically has values between 0.2--0.3, whereas for the case of BHs, the tidal deformability can be safely taken as zero for our purposes. However, a detailed description is available in Sec.~2.4 of Ref. \cite{Katerina_tidal} which is inspired by Ref. \cite{Chaves_tidal}. The tidal deformability for the BNS system can also be parametrized as an effective tidal deformability parameter $\Tilde{\Lambda}$ given as 
\begin{equation}
    \Tilde{\Lambda} = \frac{16}{13}\frac{(m_1+12m_2)m_{1}^{4}\Lambda_1 +(m_2+12m_1)m_{2}^{4}\Lambda_2}{(m_1+m_2)^{5}} \,,
\end{equation}
where subscripts 1 and 2 tag the individual companions \cite{Katerina_tidal}. The tidal deformability parameter has significant effects on the waveform and hence also the memory signal is affected. We plot a few examples to showcase the effect of tidal deformability on the memory signal in Fig.\,\ref{fig.seobnrWF}. Here we show all three possible systems NSBH, BBH and BNS, with low and high total mass (2.7 and 5 $M_{\odot}$). The oscillatory waveform is presented with a tapering at 500 Hz, while the nonlinear memory is computed from the oscillatory waveform above 200 Hz. The memory waveform is shown with and without a low-frequency cut at 10 Hz. For the BNS/BBH systems, the oscillatory signal is produced using the state-of-the-art waveform model \verb|SEOBNRv4_ROM_NRTidalv2|, while for NSBH systems \verb|SEOBNRv4_ROM_NRTidalv2_NSBH| is used. The base model for both these waveform models is the reduced order model (ROM) version of \verb|SEOBNRv4| described in Refs.~\cite{Bohe:2016gbl,Cotesta:2020qhw}. \verb|SEOBNRv4_ROM_NRTidalv2| has NR tuned tidal effects \cite{seobnr_tidal} and \verb|SEOBNRv4_ROM_NRTidalv2_NSBH| alongside tidal effects includes information about the potential tidal disruption of the NSs \cite{seobnr_tidal_nsbh,pannarale-2015}. Both of these waveform models use post-Newtonian approximation based on the effective one-body framework in the frequency domain. 

In Fig.\,\ref{fig.seobnrWF} we plot time series examples of the oscillatory and memory signals of BBH, NSBH and BNS systems. We choose two cases of masses (low $2.7 M_{\odot}$ and high $5 M_{\odot}$ ) and two cases of $\Lambda$ (low 200, high 600). The waveforms for the BNS and NSBH are tapered at the merger frequency and hence are by design only captured until the late-inspiral part of the signal, the BBH waveform is full inspiral merger and ringdown and hence the BBH memory is unsurprisingly the dominating one. It should be noted while this will always be true for the case of NSBH systems that the BBH memory will be dominating one, for BNS systems one can have memory contribution from the postmerger part which will determine the final content of BNS memory. We note that the peak memory amplitude decreases as the $\Lambda$ increases. This is true for BNS and NSBH systems regardless of the high and low mass cases. It can be attributed to the fact that with higher $\Lambda$ values, the merger frequency of the system is lowered since the size of the NS increases and hence the signal is shortened. The peak amplitude of NSBH memory is consistently higher than that of the BNS one. This is also due to the fact that the NSBH waveform is not tapered at the merger frequency while the BNS waveform is. The BNS waveform considered here does not model the postmerger phase and hence underestimates the nonlinear memory content in BNS.


\subsubsection{Memory from postmerger signal}

Another manifestation of matter effects is observable in the postmerger part of the GW signal if the remnant is an unstable/stable NS. However, this can only exist if both companions in a CBC system are NSs since for NSBH systems the remnant is always a BH. The BNS mergers can have various outcomes affecting the postmerger signal depending on the mass of the progenitors and the EOS. The cases where some postmerger signal is produced will always be the case when the merger of BNS leads to the formation of an intermediate NS \cite{Sarin_2020gxb}. If the BNS system after the merger prompt collapses to BH, then the postmerger signal is negligible. The postmerger signal is a unique signature of BNS systems and it does affect the content of the nonlinear memory. A study regarding this was conducted previously in \cite{yang-2018}. In this work, we use a vast number of publicly available NR waveforms of BNS systems and broadly survey the nonlinear memory content arising from the postmerger signal of the BNS systems as a function of different BNS parameters.

\begin{figure}[htp] 
  \centering
     \scalebox{0.3}{\includegraphics{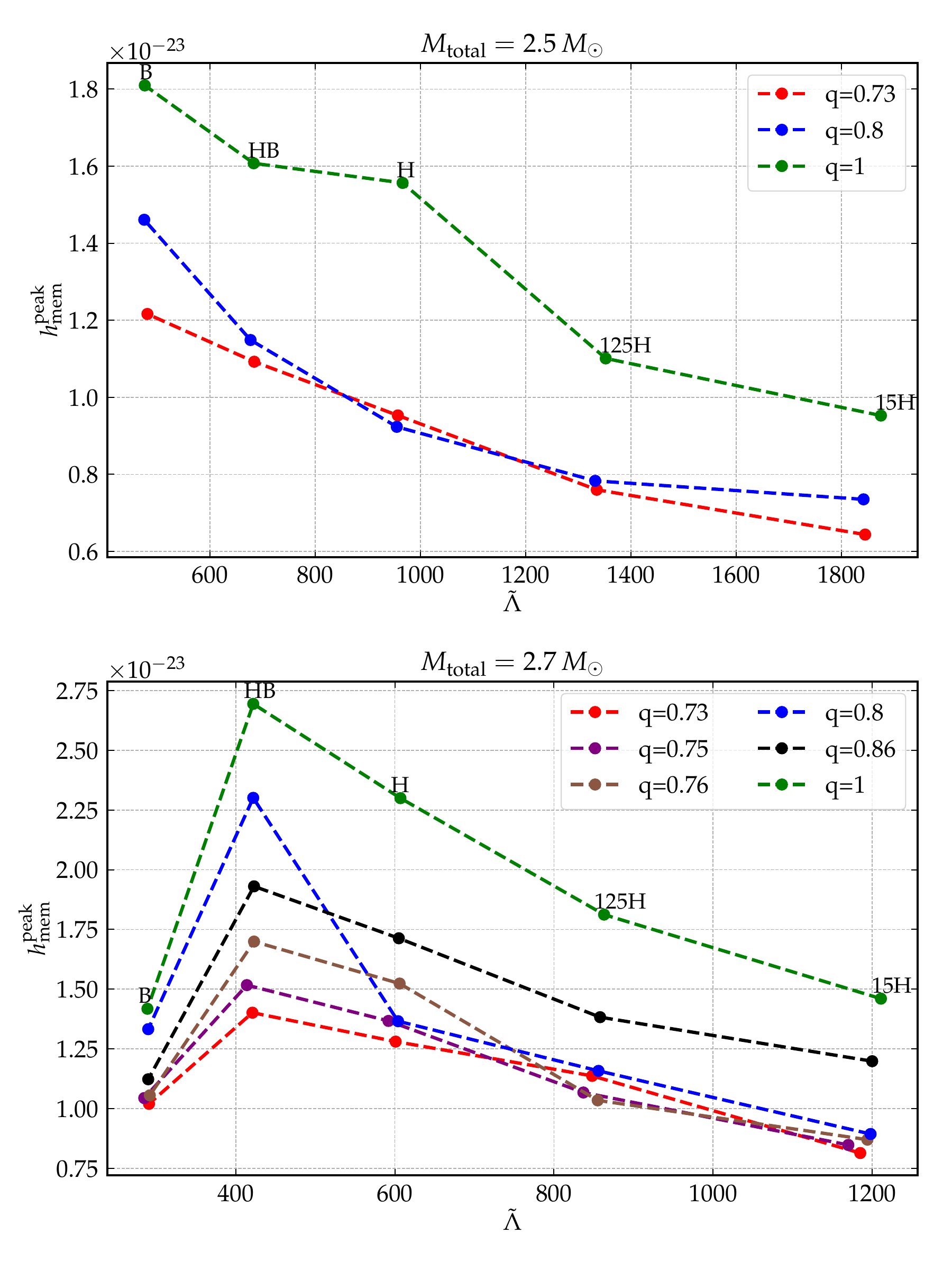}} 
\caption{The peak amplitude of the memory waveform from NR simulations as a function of its EOS for an edge-on system at 100\,Mpc are presented for different values of mass ratio. The text in each point shows the name of the EOS and the x axis corresponds to the tidal deformation parameter.}
\label{fig:hpeak_sacara}
\end{figure}

The content of nonlinear memory depends on the integrated time derivative of the \textit{parent signal} (the signal from which the nonlinear memory is produced), as can be seen in Eq.~(\ref{eq:mem_plus}). The peak displacement achieved by the memory signal is proportional to the energy emitted by the parent signal $\Delta h_{\text{peak}}^{\text{mem}} \propto \Delta E_{\text{GW}}$ and hence memory is \textit{de facto} a measure of the emitted energy during the postmerger process. This is reflected in the memory content of the postmerger signal from the BNS merger events. In Fig.\,\ref{fig:hpeak_sacara} the $\Delta h_{\text{peak}}^{\text{mem}}$ is plotted for the NR simulations SACRA presented in Ref.~\cite{SACRA}. We fix the distance to be 100 Mpc and the inclination angle to be the most favorable for the memory signal i.e., the edge-on system. Two cases of total masses are shown with different values of mass ratios. For the total mass 2.5$M_{\odot}$ a monotonic decreasing trend is observed for $\Delta h_{\text{peak}}^{\text{mem}}$ as a function of tidal deformation parameter for a system with fixed mass ratio. One can also conclude from this that the energy emitted in the postmerger phase of the neutron star is lower as the tidal deformability of the NS increases. The $\Delta h_{\text{peak}}^{\text{mem}}$ of mass ratios less than 1 are without exception lower than mass ratio 1, which again corresponds to the lower $\Delta E_{\text{GW}}$ of the postmerger signal as the mass ratio becomes different from unity.
 
For the case of total mass 2.7 $M_{\odot}$, an interesting but expected phenomenon occurs. The most compact system has a dip in the $h_{\text{peak}}^{\text{mem}}$. These systems are the ones with negligible postmerger signal as the final remnant collapses to a BH almost instantaneously. 



\section{Impact of nonlinear memory on the distinguishability of matter effects} \label{sect_3}
The GW signals from the postmerger phase of BNS systems are typically in the frequency range of about 1--4\,kHz with a duration of around a few milliseconds to more than one second, depending on the EOS and mass of the remnant. The condition for the low mass CBC event to be identified as a  system with NS unambiguously can be categorized as follows: (i) the tidal deformability parameter during the inspiral rules out zero with high confidence or (ii) there is a postmerger signal different from that of a quasi-normal-mode ringing of a BH. A caveat for this statement must be acknowledged: the so-called BH-mimickers (exotic compact objects) can show these features, but for now, we are restricting our discussion to only BH and NS as compact objects. For the case of NSBH and the BNS systems with remnants going through almost instantaneous collapse to BH, point (ii) will not be applicable as the remnant will always be a BH. 

\begin{figure}[htp] 
  \centering
    {\includegraphics[width=1\linewidth]{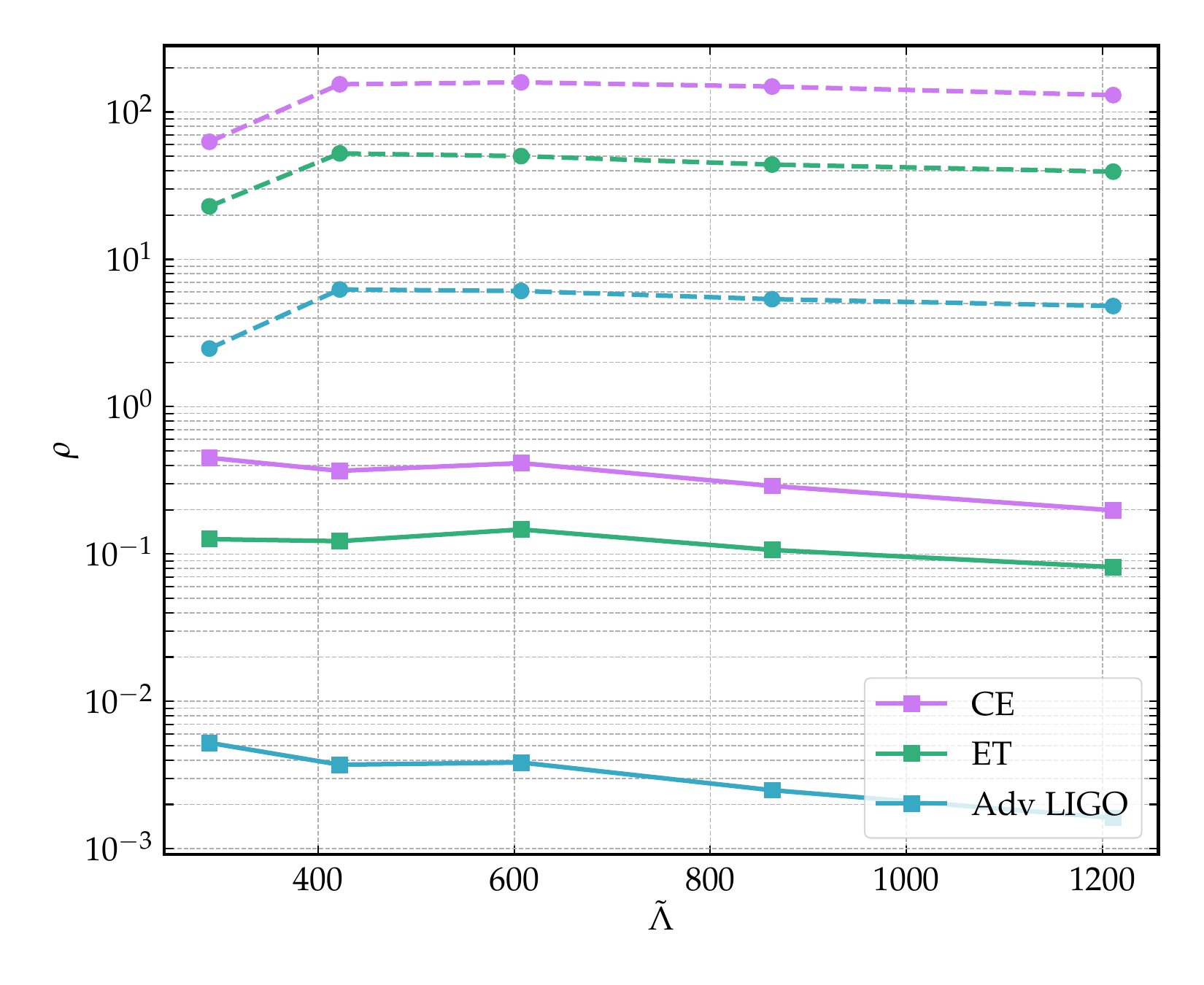}}
  \caption{SNR of postmerger phase (dashed) and nonlinear memory (solid) BNS NR waveforms for equal mass systems with $M_{\text{total}}=2.7$ at 100\,Mpc. The SNR is computed with PSDs of Advanced LIGO, ET and CE \cite{psd}.}
  \label{fig.PMsnr_sacra}
\end{figure}

Although the postmerger signal is undetected to date~\cite{PM_GW170817}, it is not a very subtle effect and thus is expected to be detected with the upcoming observing runs of the current generation of detectors or eventually with third generation detectors. 
Figure~\ref{fig.PMsnr_sacra} shows the signal-to-noise ratio (SNR) of the postmerger signal for the NR waveforms we have used in this study. The postmerger phase is defined by tapering the signal until it reaches the peak frequency (peak frequency is defined in Table~6 of Ref.~\cite{SACRA}). The inclination of the binary is chosen to optimize the nonlinear memory amplitude i.e. $\iota = \frac{\pi}{2}$. We also note that for this inclination, the detection of the gamma ray burst (GRB) jet can be difficult as the jet might not point towards the observer \cite{Metzger_2010}. One can immediately conclude that for a binary at 100\,Mpc the postmerger signal is of substantive strength to be detected unless the remnant collapses to a BH almost instantaneously, this is especially true for the current generation of detectors, i.e. Advanced LIGO at design sensitivity \cite{psd}.

We also show in Fig.~\ref{fig.PMsnr_sacra} that the SNR of nonlinear memory is around 2 orders of magnitude smaller than that of the oscillatory part of the postmerger signal. Henceforth, for the purposes of identifying the nature of the companion compact objects we will consider that the nonlinear memory will be of significance only when a postmerger signal is not present for the BNS systems.




The constraints from EM observation and astrophysical EOS inferences provide the highest NS mass limits (\textit{Tolmann–Oppenheimer–Volkoff limit}) to be  $\leq$2.0--2.5$M_{\odot}$ \cite{Nathanail_2021}. We assume that if the total mass of the binary is above 2.5$M_\odot$, the remnant is not likely to undergo postmerger oscillation and will collapse promptly to a BH. Since the precise EOS of NS is not known, the maximum mass of NS is also not known and indeed there are EOSs where BNS merger can have long-lived remnants even above 2.5$M_\odot$. We choose 2.5 $M_\odot$ as a fiducial value.


We study how nonlinear memory can play a role in identifying the nature of the object in the parameter space of systems with a total mass between 2--5$M_{\odot}$ and no postmerger signal detected. This study is an extension of the one conducted in Ref.~\cite{NSBHmem_21}, where it was shown that the nonlinear memory indeed could play a role in distinguishing between BBH and NSBH systems. 



In order to study the distinguishability of waveforms from different CBC systems, e.g. two waveforms with and without nonlinear memory, we compute the maximum correlation between them called match \cite{NSBHmem_21,match91}.
The first step is to compute the inner product between two waveforms in the frequency domain $h_1$, $h_2$ by taking into account the sensitivity of the detector defined as
\begin{equation}\label{Eqn_match}
\left\langle h_1 \mid h_2\right\rangle=4 \text{Re} \int_{-\infty}^{\infty} df \, \frac{h_1(f) h_2^*(f)}{S_n(f)}  \,,
\end{equation}
where $S_n(f)$ is the noise power spectral density (PSD) of the detector. From Eq.~(\ref{Eqn_match}), the match is computed by maximizing over coalescence time and phase as
\begin{equation}
\mathcal{M}(h_1, h_2)=\max _{t_c, \phi_c}\left[\frac{\langle h_1 | h_2 \rangle}{\sqrt{\langle h_1 | h_1 \rangle \langle h_2 | h_2 \rangle}}\right ]\,.
\end{equation}

\begin{figure*}[htp] 
    {\includegraphics[width=0.8\linewidth]{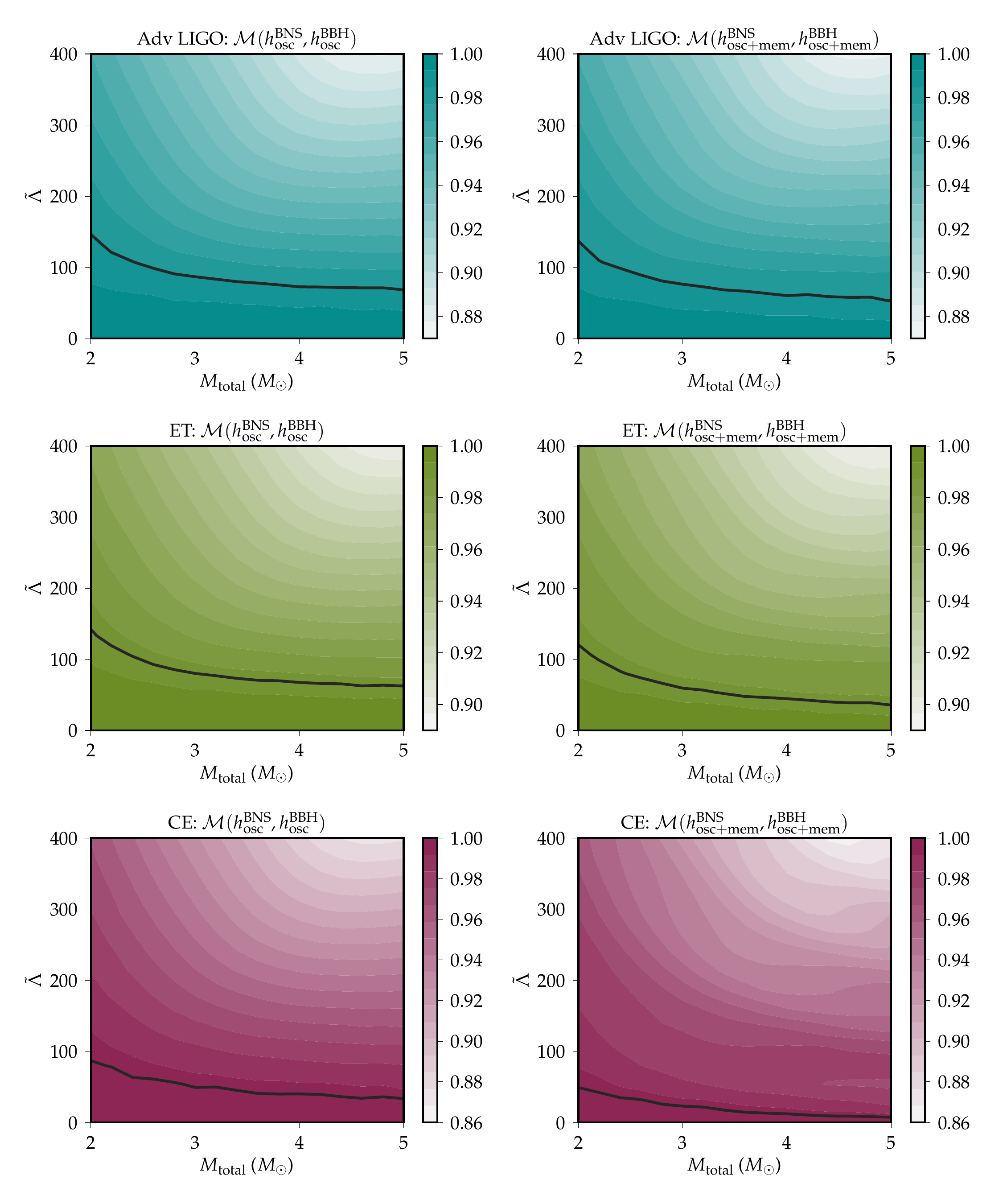}} 
\caption{Match between the GW waveform for BNS and BBH system  ($q=1$). The plot on the left side shows the match between oscillatory waveforms of GW from BNS and BBH. The right side shows the GW waveform from the same systems with the memory signal. The solid black line refers to the match of 98.5\%, 99\% and 99.5\%, at which we can distinguish the two waveforms with 90\% confidence when the source is at 20\,Mpc (SNR $\sim 14.3$), 170\,Mpc (SNR $\sim 17.4$) and 200\,Mpc (SNR $\sim 25$) for PSDs corresponding to the Advanced LIGO, ET, and CE, respectively \cite{psd}. The oscillatory waveforms are above 400\,Hz.}
\label{fig:bns_bbh}
\end{figure*}
The oscillatory part of the  BNS and BBH systems are produced using the waveform model \verb|SEOBNRv4_ROM_NRTidalv2|. The memory component is computed according to Eq.~(\ref{eq:mem_plus}) from the oscillatory part. We then perform match studies between the BBH and BNS systems for the oscillatory and oscillatory + memory waveform. The mismatches $[1-\mathcal{M}(h^{\mathrm{BNS}}, h^{\mathrm{BBH}})]$ between the oscillatory part of the signal are driven by the tidal deformability parameter in the inspiral. In contrast, the mismatches between oscillatory + memory signal also consider the difference between the memory component of BBH and BNS systems. This provides us with a metric that quantifies in what parameter space the addition of nonlinear memory impacts the signal. We perform this match study for the noise PSDs corresponding to the Advanced LIGO, Einstein Telescope (ET), and Cosmic Explorer (CE) \cite{ligo_2015,et_science,cs_science}. 

The total mass of binaries between 2--5$M_{\odot}$ with $q=1$ and tidal deformability of individual NSs in the binaries ranging from 0--400 ($\Lambda = 0 $ corresponds to BBH). This $\Lambda$ range intends that almost above 400, the oscillatory signal will be more dominant in identifying the nature of the companion objects, so we consider it only until 400 for nonlinear memory purposes. The oscillatory waveform is generated starting from 400\, Hz and the nonlinear memory component is computed from 200\,Hz to allow for a longer time of integration to get rid of artifacts (the effect on nonlinear memory signal strength is negligible).
After the generation of the memory signal, we apply a high pass filter (low-frequency cut) to the memory waveform at 10 \,Hz, this high pass filter makes the nonlinear memory a burst-like signal centered around the peak amplitude of the oscillatory signal.

In Fig.\,\ref{fig:bns_bbh}, we show the results for the matches between BNS and BBH systems with and without memory. The confidence region of distinguishability between two waveforms is computed from the match as in Ref.~\cite{match_confid},
\begin{equation}
    \mathcal{M}(h_1,h_2) \geq 1-\frac{\chi_{k}^{2}(1-p)}{2\rho^{2}} \,,
    \label{eqn:match}
\end{equation} 
where $\chi_{k}^{2}(1-p)$ is the normal chi-squared distribution with $k$ degrees of freedom for $(1-p)$ probability (here $k=3$ from considering the parameters $\tilde{\Lambda}$, $M_{\text{total}}$ and $q$) and $\rho$ is the SNR. The solid black line in the plot divides the parameter space with the distinguishability of the system at 90\% confidence (the parameter space above the black line is distinguishable). The SNR is computed for the CBC systems at different distances as 20 Mpc for LIGO, 170 Mpc for ET, and 200 Mpc for CE with 3 degrees of freedom (total mass, mass ratio and tidal deformability of the NS). This is done to account for the vast difference the SNR will have in different detectors when the source is at the same distance. Using Eq.~(\ref{eqn:match}), the distinguishability contour can be computed for any signal strength (SNR or distance). Since our purpose here was to study the impact of nonlinear memory, we chose these different values of distance (SNR) to have similar distinguishability of BNS and BBH systems using only oscillatory part for the three detectors considered; i.e. we choose closer distance systems for LIGO, and farther for ET and CE. 

The impact of nonlinear memory in distinguishing BNS and BBH systems is less significant than it was for the NSBH case \cite{NSBHmem_21}. Still, there is an overall enlargement of the parameter space where one can distinguish these signals. This improvement can be especially seen for CE, which was expected since its PSD allows for the best measurement of nonlinear memory.

We have conducted the same study for the BNS and NSBH binaries. These details are presented in  Appendix~\ref{app:nsbh} where we find again that the enlargement of distinguishable parameter space with the addition of nonlinear memory is not significant for Advanced LIGO. However, for Cosmic Explorer the improvement is substantial. 

For completeness, we also explore the possibility that nonlinear memory from the postmerger NS signal by itself can distinguish between various EOSs. We conducted a mismatch study between only the nonlinear memory waveform from different EOSs for postmerger BNS systems with equal mass binaries and total mass 2.7 $M_{\odot}$. The results are reported in terms of mismatches (1-match) in Table~\ref{tab:Mismatch_nr} in percentages. The memory waveform from the BNS system with prompt collapse to a BH almost behaves similarly to the waveform from the BBH merger of the same masses. The rest of the waveforms show a mismatch from BBH in the range of 2.5 \% -- 5 \%. This can be indeed interesting but as already discussed previously, for these events the postmerger oscillatory signal will be much stronger than the nonlinear memory signal. 

\begin{table}[htp]
\centering
\begin{tabular}{c c c c c c c}
\hline
 \hline
 EOS & B & HB   & H   & 125H & 15H   & BBH \\
 \hline \hline
B    & 0 & 2.2 & 1.63 & 2.16 & 2.88 & 0.39 \\ 
HB   &   & 0   & 0.15 & 0.11 & 0.2  & 3.57   \\
H    &   &     & 0    & 0.11 & 0.38 & 2.76 \\ 
125H &   &     &      & 0    & 0.12 & 3.35   \\
15H  &   &     &      &      & 0    & 4.32  \\ 
 \hline
\end{tabular}
\caption{The mismatch between postmerger BNS nonlinear memory waveforms from NR simulations with $q=1$ and $M_{\text{total}} = 2.7 M_{\odot}$ for various EOSs; also, we have included the BBH signal in this analysis as well for comparison. The mismatches are computed with a low frequency cut at 10\,Hz for Cosmic Explorer PSD. The values are given in percentages.}
\label{tab:Mismatch_nr}
\end{table}

\section{Detection prospects of postmerger memory from stacking multiple events} \label{sect_4}
Nonlinear memory is a subdominant effect and hence it is challenging to detect it in individual GW events. Some studies have shown that the nonlinear memory of individual events may not be loud enough to be detected by the current generation of ground based detectors and hence stacking multiple events will be needed \cite{lasky_GW150914mem,mem_patricia,Review_Grant2022,yang-2018}. 
The method employed in these studies is the so-called SNR stacking. This method aims to constructively sum the signal energy, parametrized in this case by the SNR, from a population of events. The cumulative SNR $\rho_{\text{eff}}$ of $N$ events, each with individual SNR $\rho_i$ is given as

\begin{equation} 
    \rho_{\mathrm{eff}} = \sqrt{\sum_{N=1}^{i}\rho_{i}^{2}}\,.
    \label{eqn:snr_sum}
\end{equation} 


In this section, we focus on the prospects of detecting postmerger NS nonlinear memory from a population of BNS merger detections. 
To achieve this, we use the NR waveforms previously described in Sec.~\ref{sect_2}. We pick the most optimistic case with regards to the nonlinear memory amplitude of total mass and mass ratio, i.e., total mass $=2.7 M_{\odot}$ and mass-ratio unity. We chose three EOSs that can sufficiently probe the parameter space for our purposes: B, HB and 15H. EOS B corresponds to the near instantaneous collapse of the remnant to a BH. EOSs HB and 15 H give the highest and lowest amplitude of nonlinear memory with a long-lived remnant. 

We generate the events with fixed intrinsic parameters as mentioned above. For the extrinsic parameters, we fix the inclination angle of the binary as optimal for the nonlinear memory i.e., edge-on. We consider the redshift distribution, where the events are sampled uniformly between 0 and a maximum redshift of $z_{\mathrm{max}}=0.1$. Here we are investigating a subtle effect that will only be detectable in a population of events with a very high SNR. The low SNR of memory for the high redshift events will take a random walk in Gaussian noise parameter space since the high redshift events lead to further diminished memory SNR. Therefore, we focus on a nearby universe with a maximum redshift of $z_{\mathrm{max}}=0.1$, where events can be reasonably regarded as uniformly distributed in volume \cite{Review_Grant2022}.
The maximum luminosity distance ($D_L$) is computed from the redshift as in Ref.~\cite{hor_dist}. 
Then for each case of EOS, we generate ten events to compute the effective cumulative SNR as given in Eq.~(\ref{eqn:snr_sum}). We repeat this process 50 times to obtain a distribution in cumulative SNR for each case of EOS. The results are reported in terms of the median and the 25\% spread from the median of cumulative SNR. We follow the same procedure for 100 events.

\begin{figure}[htp] 
  \centering
    {\includegraphics[width=0.8\linewidth]{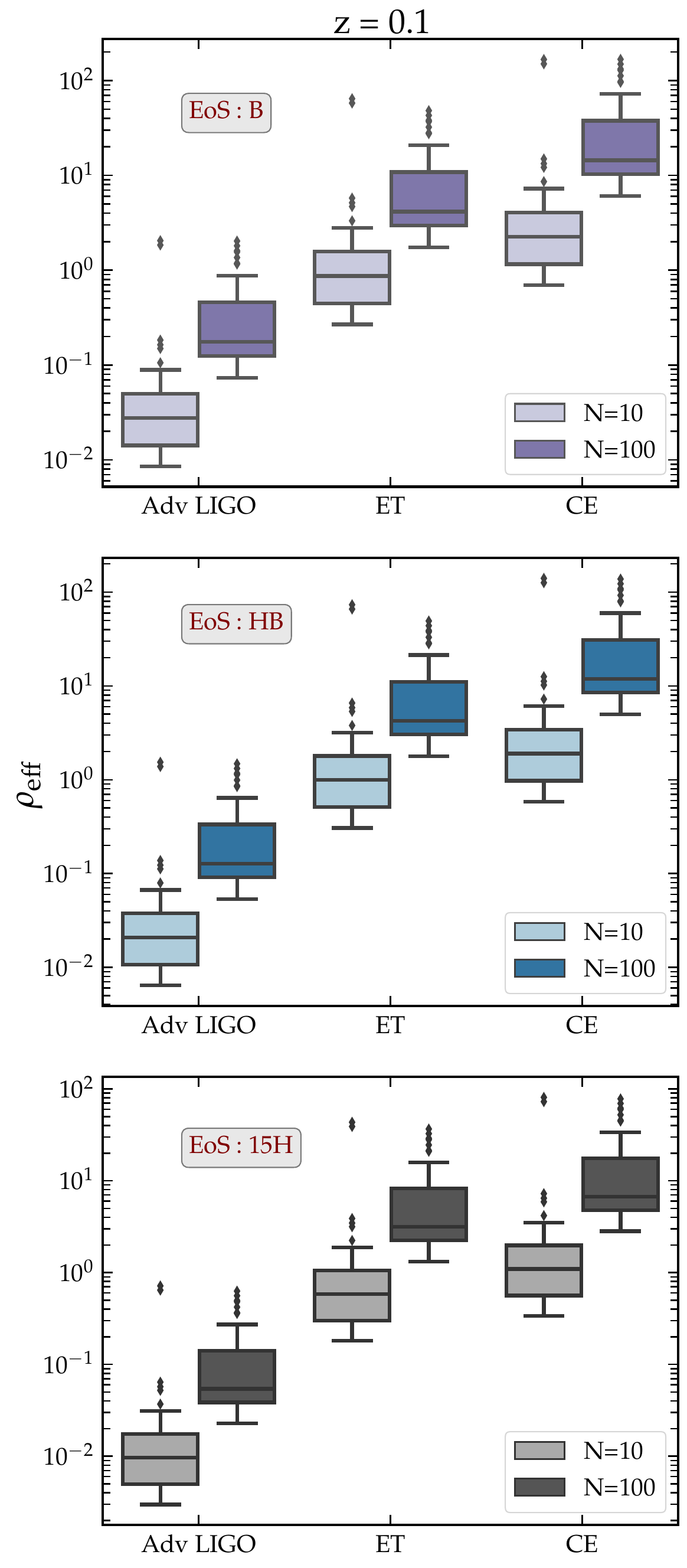}} 
\caption{Cumulative memory SNR of 10 and 100 events corresponds to Advanced LIGO, Einstein Telescope, and Cosmic Explorer design sensitivity. The plot shows the events within the redshift of $z=0.1$. The three rows correspond to events with the NR waveform template of EOSs B, HB and 15H. Each box plot corresponds to 50 realizations. The solid black line indicates the median value and the edges correspond to the spread of cumulative SNR values that fall within 25\% of the median.}
\label{fig:snr_eff}
\end{figure}

In Fig.\,\ref{fig:snr_eff}, we show the cumulative SNR for 10 and 100 detected events within maximum redshift of 0.1 using the noise PSDs of Advanced LIGO, Einstein Telescope and Cosmic Explorer \cite{psd}.
The effective SNR of the memory waveform is computed for the randomly chosen 10 and 100 events of varying distance within the maximum luminosity distance \cite{hor_dist} for the design sensitivity of Advanced LIGO, ET and CE. The analysis is repeated 50 times to get enough statistics. The solid black line on each box plot corresponds to the median of cumulative SNR for 50 realizations. We compute the cumulative SNR of nonlinear memory for different EOSs (B, HB and 15H) with increasing order of tidal deformability. For all three sets of waveforms, the $\rho_{\text{eff}}^{\text{median}}$ is below one with the sensitivity of Advanced LIGO for a population of 100 events, whereas $\rho_{\text{eff}}^{\text{median}} \gtrsim 4$ and $\rho_{\text{eff}}^{\text{median}} \gtrsim 8$ with ET and CE  sensitivity, respectively. The possibility of detecting nonlinear memory from the postmerger signal is very low even with the third generation detectors if we have only ten events since the $\rho_{\text{eff}}^{\text{median}}$ reduced to less than two. The emulation of the binary neutron star (BNS) merger occurrences in this study is primarily influenced by the local distribution of events rather than the overall universe. Our objective is to assess the possibility of detecting gravitational wave memory from the BNS population by estimating the likelihood of high event rates in the nearby universe.




In Appendix~\ref{app:bbh_eff}, we show the cumulative SNR of nonlinear memory from BBH events with the same procedure as discussed in this section. We also investigated the possibility of distinguishing different EOS and BBH mergers using nonlinear memory. Table~\ref{tab:snr_diff} shows the difference between the $\rho_{\text{eff}}^{\text{median}}$ for the three EOSs shown in Fig.\,\ref{fig:snr_eff} in addition, we also included the BBH events discussed in Appendix~\ref{app:bbh_eff}.
The $\Delta \rho_{\text{eff}}^{\text{median}}$ for the BNS systems with different remnants, one which directly collapses to a BH and the other forming a postmerger NS, show a difference in cumulative SNR of greater than 5, whereas the nonlinear memory from BBH and BNS differs by more than 100.

\begin{table}[htp]
\centering
\begin{tabular}{c c c c c}
\hline
 \hline
 EOS & B  & HB  & 15H  & BBH \\
 \hline \hline
B    & 0 & 2.3 & 7.48 & 124  \\ 
HB   &   & 0   & 5.18 & 126  \\
15H  &   &     & 0    & 132  \\ 
 \hline
\end{tabular}
\caption{Difference between the median of $\rho_{\text{eff}}$ for the population of 100 events distributed within $z<0.1$ with the CE sensitivity.}
\label{tab:snr_diff}
\end{table}

In the four compact binary systems considered for the study, the median effective SNR of GW memory from the BBH system has an order of magnitude improvement compared to BNS postmerger memory. The BNS merger rate inferred by the GWTC-3 catalog is between 10 -- 1700 $\mathrm{Gpc}^{-3}\mathrm{yr}^{-1}$ \cite{GWTC3_rates}. 
Considering a fiducial rate of $1000\,\mathrm{Gpc}^{-3}\mathrm{yr}^{-1}$, 0.9 years of observing time are required to get the effective SNR from 100 BNS mergers within $z<0.1$ with CE. Also, the SNR falls with an increase in maximum redshift. 
\section{Linear memory of ejecta from binary neutron star and black hole-neutron star systems}\label{sect_5}

In BNS and NSBH systems, a substantial amount of matter can be dynamically ejected depending on the configuration of the system. This can lead to another nonoscillatory emission of GWs, the so-called linear memory. This contribution to the GW signal is usually present if the interacting particles in a GW source are or become unbound. An example is binaries on hyperbolic orbits~\cite{turner_will-1978,deVittori-2014}, but linear memory has also been studied e.g for gamma-ray burst jets~\cite{sago-2004,birnholtz-2013,yu-2020} and in the context of supernova explosions and their associated asymmetric mass loss due to neutrino emissions~\cite{turner-1978,burrows-1995,kotake-2005,ott-2008,mukhopadhyay-2021}. A general formula for the linear memory from a system of $N$ bodies with changing masses $m_A$ or velocities $\bm v_A$ is given in Ref.~\cite{thorne-1992}, the change to the metric perturbation in TT gauge is
\begin{align}
    \Delta h_{jk}^\mathrm{TT} = \Delta \sum_{A=1}^{N} \frac{4 m_A}{r \sqrt{1-v_A^2}} \left[\frac{v^j_A v^k_A}{1-\bm{v}_A \cdot \bm{N}}\right]^\mathrm{TT}\,,
\end{align}
where the $\Delta$ means to take the initial and final value of the summation, $\bm N$ denotes a unit vector pointing from the source to the observer and $r$ is the distance to the source. The particles composing the system of masses $m_A$ moving with velocities $\bm v_A$ in our case might refer to the individual pieces of matter becoming unbound in a BNS/NSBH coalescence. However, modeling the ejecta dynamics is a complex task as it depends on most of the binary's parameters like mass ratio, equation of state, spins, composition or magnetic fields. Nonetheless, we want to provide some estimates to show that the nonlinear memory signal we are investigating in this paper is not noticeably contaminated by a potential linear memory signal.

In order to estimate the maximum amount of linear memory created by dynamical ejecta, we assume that the ejected mass $m_\mathrm{ej}$ travels radially away from the much larger remnant mass with velocity $v_\mathrm{ej}$. In this case the amplitude of the linear memory created can be approximated by~\cite{braginskii-1987}
\begin{align}
    \Delta h = \frac{2 G \, m_\mathrm{ej} \, v_\mathrm{ej}^2}{c^4 \, r} \,.
    \label{eq:linmemapprox}
\end{align}
This is the model creating the maximum linear memory amplitude, whereas a perfectly spherically symmetric outflow of the ejected mass would not lead to any linear memory. The emission direction is mostly perpendicular to the direction of the ejecta, thus if the material is ejected in the orbital plane, the maximum amplitude of the linear memory can be observed in the $z$ direction.

From studies about dynamical ejecta in NR simulations of a wide range of BNS systems~\cite{dietrich-2016,radice-2016,radice-2018,hotokezaka-2012}, we find that the ejected mass is typically below $0.01 M_\odot$ but can go up to $0.1 M_\odot$ for certain EOSs, more unequal mass systems or eccentric mergers. The velocity of the ejecta is usually about $0.2 c$ with values up to $0.4 c$, mostly directed in the orbital plane with the velocity component perpendicular to the orbital plane generally around or well below $0.1 c$, with equal mass BNSs commonly having more ejecta out of the orbital plane. The azimuthal opening angle of the ejecta is often between $\sim \pi$ and $2 \pi$ with more unequal mass systems on the lower and equal mass BNS with more uniform radial ejecta. According to Eq.~(\ref{eq:linmemapprox}), the order of magnitude of the linear memory can be estimated as 
\begin{align}
    \Delta h = 3.8 \times 10^{-25} \left(\frac{m_\mathrm{ej}}{0.01 M_\odot}\right) \left(\frac{v_\mathrm{ej}}{0.2 c}\right)^2 \left(\frac{r}{100\,\mathrm{Mpc}}\right)^{-1} \,,
    \label{eq:linmemestimate}
\end{align}
which is about 2 orders of magnitude below the nonlinear memory from the coalescence. Summing up, the most linear memory amplitude is expected when looking face-on at an unequal mass BNS system. On the other hand, due to the symmetric ejection in the orbital plane, almost no linear memory is expected for equal mass BNS perpendicular to the orbital plane. However, they can show small amplitudes emitted in the orbital plane.

Potential dynamical ejecta have also been studied for NSBH mergers~\cite{kyutoku-2013,kyutoku-2015,foucart-2016}. Depending on the binary parameters and the EOS, the neutron star can either be tidally disrupted or plunge as a whole into the black hole~\cite{pannarale-2015}. In the latter case, the mass ejection is almost negligible, whereas a substantial amount of matter, typically about 0.01--0.1 $M_\odot$, can be ejected with velocities $v_\mathrm{ej} \sim 0.2 c$ in tidal disruption events. The amount of matter ejected is generally larger for stiff EOSs and for large black hole spins. The ejection of the matter is predicted mostly inside the orbital plane with an opening angle of $\sim \pi$, therefore like for BNS, linear memory is expected mostly perpendicular to the orbital plane with a similar amplitude if the neutron star is tidally disrupted. 

In order to estimate the importance of the linear memory, not only its final amplitude has to be considered, but also how fast it rises to the final value as that determines the frequency content of the memory burst and thus if the signal is observable at all. The time evolution of the linear memory waveform from the ejecta can be phenomenologically modeled as $h (t) = \Delta h (1+e^{-t/\tau})^{-1}$, where $\Delta h$ is the final amplitude that can be estimated according to Eq.~(\ref{eq:linmemestimate}) and $\tau$ denotes the duration of the ejection process~\cite{yang-2018}. The corresponding frequency domain waveform is $h(f) \sim i \pi \tau /\sinh{(2\pi^2 f \tau)}$, thus one can directly see that larger values of $\tau$ correspond to emission at lower frequencies, which is more difficult to detect. In order for linear memory to be emitted at detectable frequencies, the characteristic timescale of the ejection process has to satisfy $\tau < $~few~ms. However, even then the linear memory in its most favorable case is of about an order of magnitude smaller than the nonlinear memory.

In BNS and NSBH coalescences, linear memory is not only expected to arise from dynamical ejecta but also from potential gamma ray bursts (GRBs) associated with these sources~\cite{sago-2004,birnholtz-2013}. GRBs involve explosions which release huge amounts of energy $E_j$ in accelerating jets (a forward jet and a backward jet), reaching
ultrarelativistic velocities. The amplitude of the linear memory from a GRB scales with
\begin{align}
    \Delta h = \frac{2 E_j \beta^2}{r} \frac{\sin^2 \theta}{1 - \beta \cos \theta}\,,
\end{align}
where $\beta \sim 1$ is the velocity of the jet divided by the speed of light and $\theta$ is the angle between the jet direction and the line of sight from the source to the observer. The linear memory is antibeamed, with no emission in jet direction but otherwise almost isotropic emission. 
For GRBs, the acceleration time of the jet and thus the rise time of the linear memory is estimated to be between 0.2 and 3 ms, depending on the opening angle and composition of the jet. This results in a GW memory burst well within the frequency band of aLIGO. However, the amplitude is about 2 orders of magnitude too small to collect significant SNR, though large uncertainties exist. 


\section{Discussions}\label{sect_6}
\label{sec:discussions}
In this work, we have investigated various aspects of gravitational wave memory from binary neutron star mergers. We find that nonlinear memory of the GW signal can contribute in distinguishing BNS systems from other compact binary systems. This effect is quantified in terms of the mismatch between waveforms from different systems. We have shown that adding nonlinear memory to the waveform models increases the parameter space for the distinguishability of equal mass BNS systems from the same mass BBH and NSBH systems. The distinguishability between various compact systems is more apparent with the third generation detector sensitivity for a range of parameter space in the lower mass gap.

We also analyzed the possibility of detecting the nonlinear GW memory effect from the postmerger phase of BNS systems using numerical relativity simulations. We find that the SNR of the memory signal from an individual event is almost 2 orders of magnitude lower than the oscillatory waveform, irrespective of the EOS. Hence, instead of finding the SNR of the memory waveform from a single event, we followed the approach of finding cumulative memory SNR from a population of BNS events for various EOSs. We consider the  NR simulations of different EOSs and postmerger fates. For the BNS events populated within the redshift of $z=0.1$, it requires at least a population of 100 events to have a confident detection of nonlinear memory by the future ground based detectors. The chance of detecting nonlinear memory from a population of 100 BNS systems would be less feasible with the current ground based detector. However, with ET and CE, the median of effective SNR  over different realizations for a population of 100 events are above  4 and 8, respectively. These SNR values for the population of BNS systems with various EOS almost lie in the same range, with a difference in effective SNR of around 5. This difference of stacked SNR is not enough to distinguish between various EOS studied here. We found an increase in effective SNR  by a factor of 10 for the population of BBH events compared to BNS events with the current and third generation detector sensitivity. Thus, in the absence of EM observation and tidal deformability measurement or when the GWs from the merger/postmerger remnant are below the detection sensitivity, the nonlinear memory from a population of events can help to distinguish between BNS and BBH systems. We have also shown that the effect of linear memory is negligible as compared to that of nonlinear memory in BNS systems.

\begin{acknowledgments}
The authors would like to thank Maria Haney and Giovanni Prodi for their discussions and encouragement throughout the duration of this work. 
S.T. is supported by the Swiss National Science Foundation (SNSF) Ambizione Grant No. PZ00P2-202204. The authors acknowledge the support of the University of Zurich for the provision of computing resources. 


\end{acknowledgments}

\appendix

\renewcommand{\thesubsection}{\Alph{subsection}}

\section{$\Delta h^{\text{peak}}_{\text{mem}}$ from CoRe database NR waveforms}\label{app:coreDB}

\begin{figure}[htp] 
    {\includegraphics[width=1\linewidth]{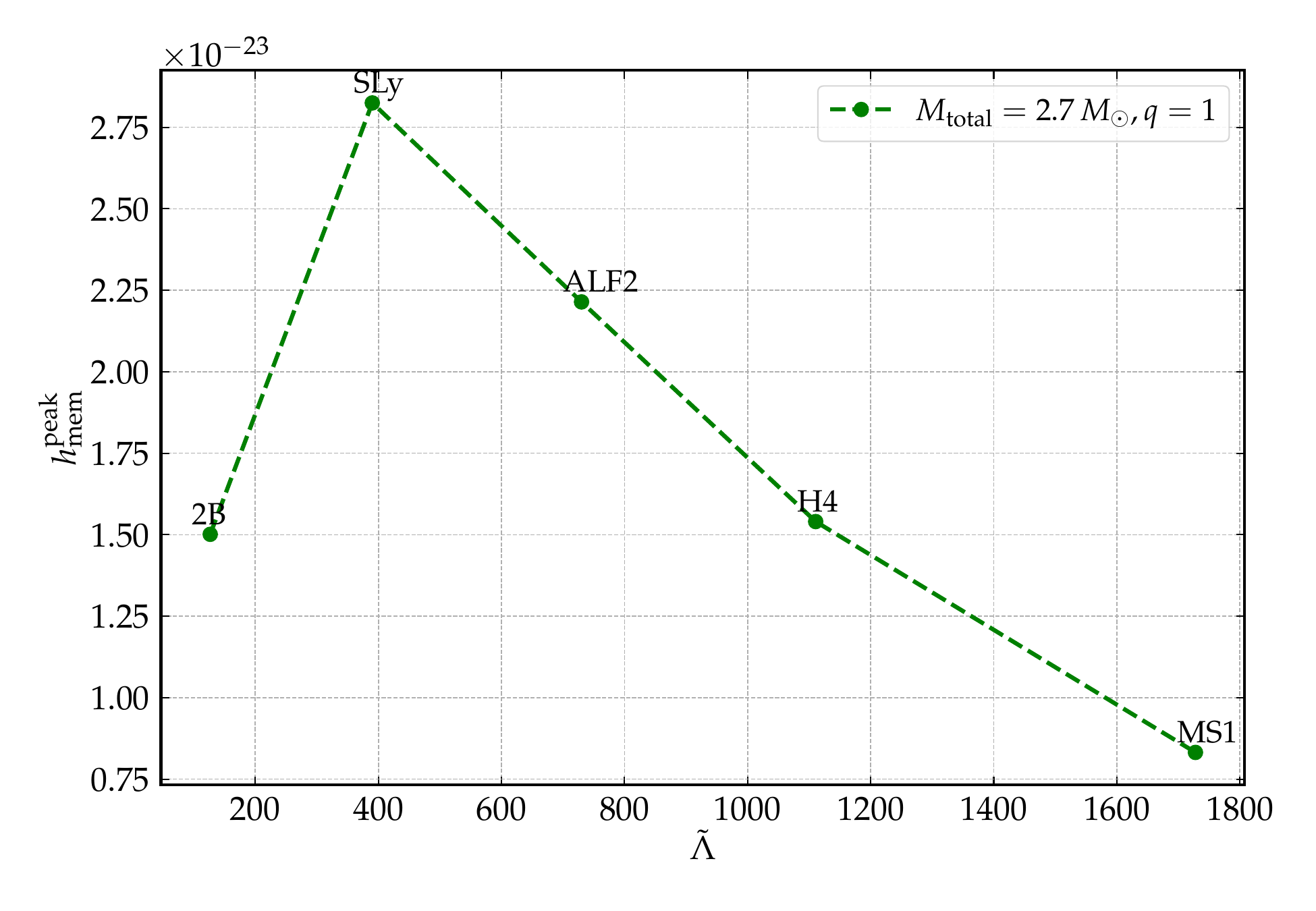}}
\caption{Peak amplitude of the nonlinear memory waveform as a function of its EOS for the CoreDB waveforms. All the events are with a mass ratio of 1 and a total mass equal to $2.7 M_{\odot}$. The source is fixed at 100\,Mpc with an orbital inclination angle of $90^{\circ}$. The text in each point shows the name of EOS and the x axis corresponds to the tidal deformation parameter.} 
\label{fig:coreDB}
\end{figure}

We conducted the study of computing nonlinear memory using the numerical relativity BNS waveforms from the CoRe database \cite{CoRe_2018}. Figure.\,\ref{fig:coreDB} shows the $\Delta h^{\text{peak}}_{\text{mem}}$ with increasing value of tidal deformability for a set of NR waveforms with equal mass binaries and 2.7 $M_{\odot}$ total mass. We fix the luminosity distance and inclination angle at 100\,Mpc and $90^{\circ}$. The waveforms are selected based on the highest resolution available for a given EOS in the range of (96,256) with tidal deformability values ranging from 100--1800 \cite{CoRe_2018}. Only a selected number of EOSs with $q=1$ and $M_{\text{total}}=2.7 M_{\odot}$ are  shown here. For a window of tidal deformability, we choose only one EOS. We find the same trend of $\Delta h^{\text{peak}}_{\text{mem}}$ value with an increase in tidal deformability for CoRe database NR waveforms as in SACRA NR simulations with the more compact BNS system directly collapsing to the black hole. 

\section{BNS-NSBH match}\label{app:nsbh}
\begin{figure*}[!t] 
    {\includegraphics[width=0.8\linewidth]{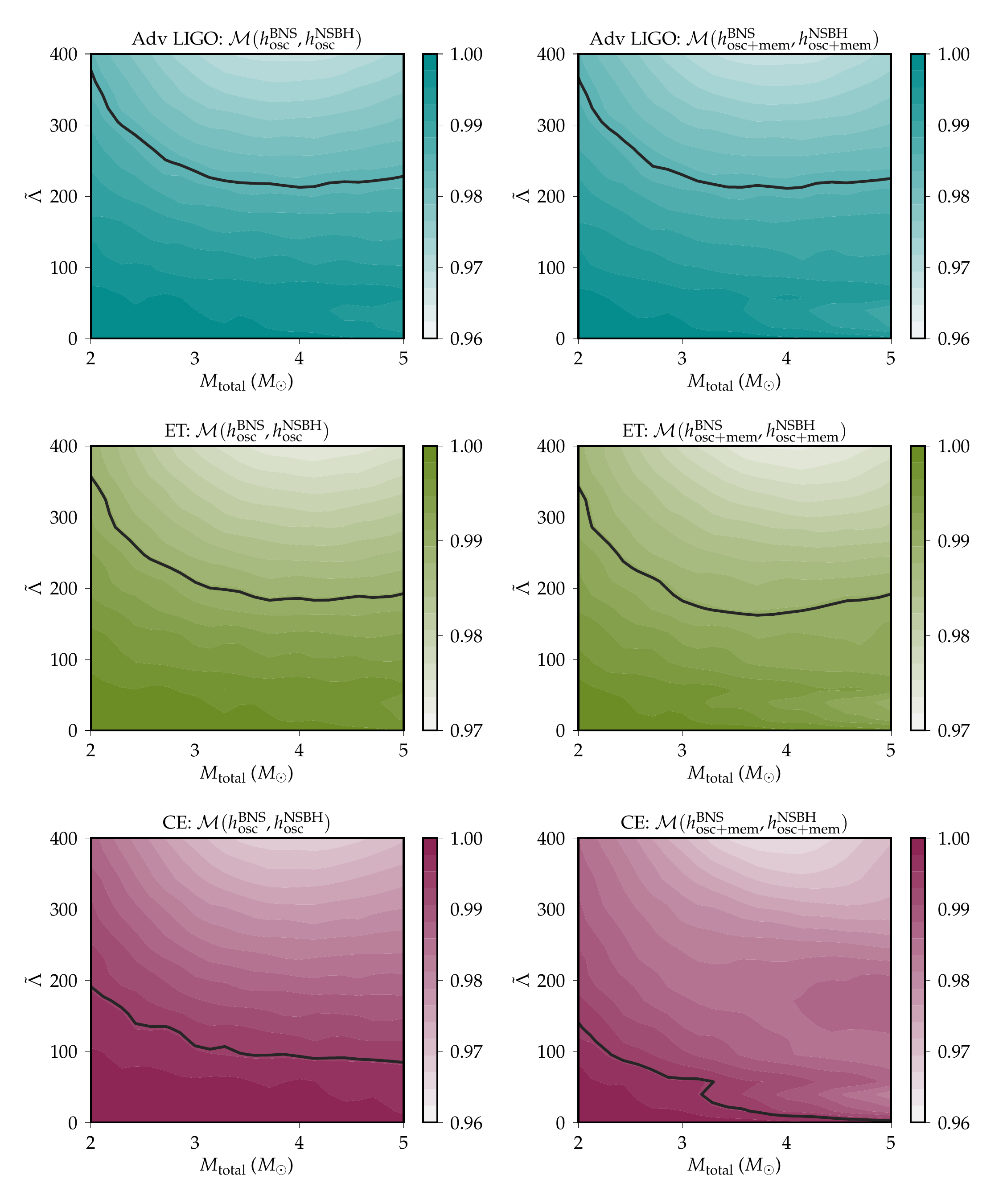}} 
\caption{Match between the GW waveform for the BNS and the NSBH system ($q=1$). The plot on the left side shows the match between oscillatory waveforms of GW from BNS and NSBH. The right side shows the GW waveform from the same systems with the memory signal. The solid black line refers to the match of 98.5\%, 99\% and 99.5\%, at which we can distinguish the two waveforms with 90\% confidence when the source is at 20\,Mpc (SNR $\sim 14.3$), 170\,Mpc (SNR $\sim 17.4$) and 200\,Mpc (SNR $\sim 25$) for PSDs corresponding to the Advanced LIGO, ET, and CE, respectively. The oscillatory waveforms are above 400\,Hz.} 
\label{fig:bns_nsbh}
\end{figure*}

Here we study the match between BNS and NSBH waveforms with and without the memory, which is shown in Fig.\,\ref{fig:bns_nsbh}. The BNS and NSBH waveforms are generated with \verb|SEOBNRv4_ROM_NRTidalv2| and \verb|SEOBNRv4_ROM_NRTidalv2_NSBH| models, respectively. It exhibits the same trend as in Fig.\,\ref{fig:bns_bbh}, such that adding nonlinear memory to the oscillatory waveform does not show significant distinguishability between BNS and NSBH systems for the Advanced LIGO and ET detectors. In the case of CE, adding memory increases the mismatch between the two systems. The solid black line in Fig.\,\ref{fig:bns_nsbh} indicates the region above which the two waveforms can distinguish with 90\% confidence for CBC systems at the same distance as mentioned in Sec.\,\ref{sect_3}. However, in the previous study \cite{NSBHmem_21}, it showed that the  addition of nonlinear memory led to enlarging the parameter space of distinguishability between NSBH and BBH systems for the current and future detectors, with more prominently visible in the case of CE detector. 


\section{Memory from binary black hole mergers}\label{app:bbh_eff}

\begin{figure}[htp] 
  \centering
    {\includegraphics[width=0.8\linewidth]{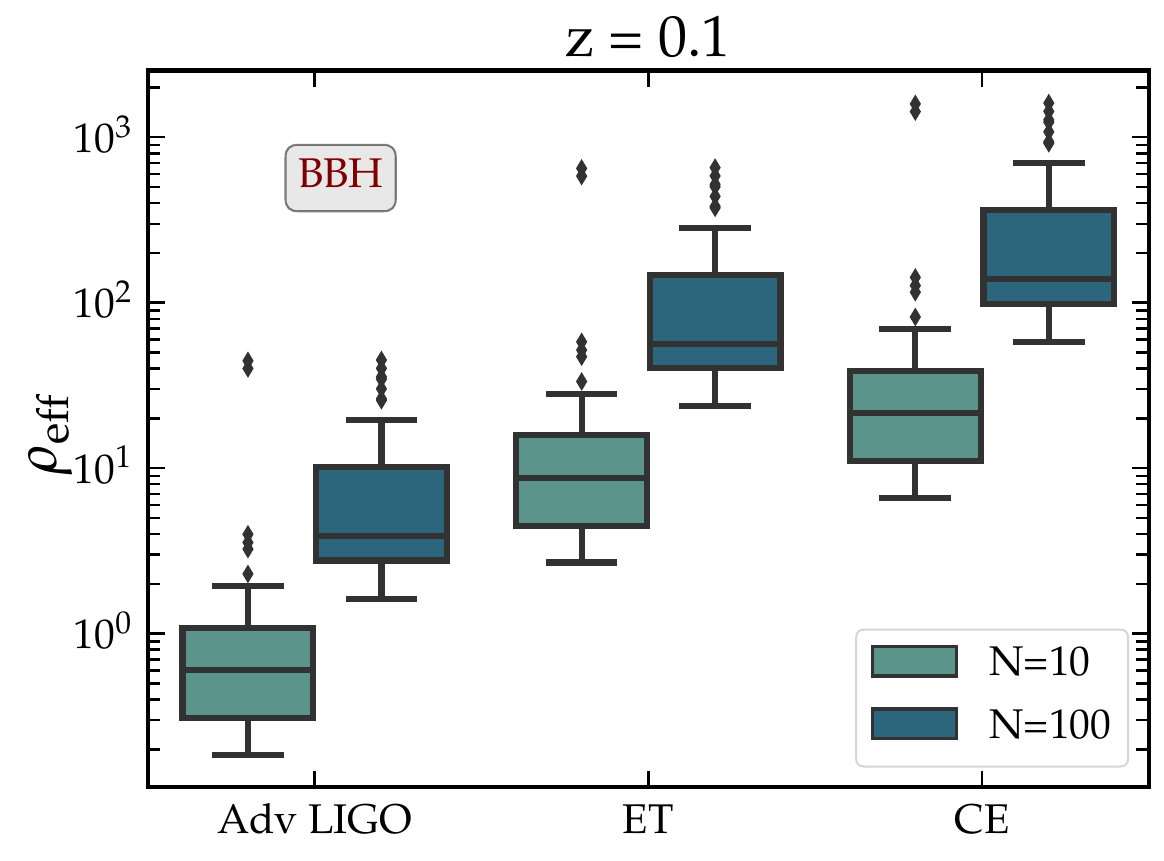}} 
\caption{The cumulative memory SNR of BBH waveforms with the $\mathtt{SEOBNRv4\_ROM\_NRTidalv2}$ model (i.e., $\Lambda_{1}=\Lambda_{2}=0$). All the events with mass ratio 1 and total mass equal $2.7 M_{\odot}$. The plot shows the distribution of 10 and 100 events within the maximum redshift of $z=0.1$. The solid black line corresponds to the median of the cumulative SNR from 50 realizations.}
\label{fig:snrEFF_Bbh}
\end{figure}

As discussed in Sec.\,\ref{sect_4}, the SNR of nonlinear memory for CBC systems is almost an order of magnitude lower than the oscillatory signals. The cumulative SNR of nonlinear memory from the BBH system was computed with the same procedure as in Sec.\,\ref{sect_4} for the BNS systems. Figure \ref{fig:snrEFF_Bbh} shows the cumulative SNR for BBH edge-on system memory waveforms with events distributed within the maximum redshift of $z=0.1$. Comparing to Fig.\,\ref{fig:snr_eff}, we can see that for the BBH system with the same component masses as BNS systems, the cumulative SNR of nonlinear memory is almost an order of magnitude greater than BNS NR waveforms for PSDs corresponding to all three detectors: Advanced LIGO, ET and CE. Thus, the nonspinning 100 BBH events with total mass lie at $2.7 M_{\odot}$ and orthogonally aligned orbit to the detector's line of sight, populated within maximum redshift of $z=0.1$. The median effective SNR for 100 events, averaged over 50 realizations, is approximately 4 for Advanced LIGO. However, the SNR significantly increases to around 60 for the ET and exceeds 130 for the CE. The third generation detectors like ET and CE can detect memory with a cumulative SNR greater than 8, even for ten events. 

The observed mass distribution for BBH events shows peaks at $10 M_{\odot}$ and $35 M_{\odot}$ \cite{GWTC3_rates}, as mentioned in Sec.\,\ref{sect_2} the increase in black hole mass increases the amplitude of nonlinear memory. Hence, the probability of memory detection from BBH events is more likely with the upcoming observing runs of Advanced LIGO.

\clearpage
\nopagebreak
\bibliography{ref}

\end{document}